# Epistemic Complexity and
# the Journeyman-Expert Transition

## Thomas J. Bing[1] and Edward F. Redish[2]

*[1]Department of Physics, Emory University, Atlanta, Georgia 30322, USA*
*[2]Depts. of Physics and Curriculum & Instruction, University of Maryland, College Park, MD 20742, USA*

**Abstract.** Physics students can encounter difficulties in physics problem solving as a result of failing to use knowledge that they have but do not perceive as relevant or appropriate. In previous work the authors have demonstrated that some of these difficulties may be epistemological. Students may limit the kinds of knowledge that they use. For example, they may use formal manipulations and ignore physical sense making or vice versa. Both beginning (novice) and inter-mediate (journeymen) students demonstrate these difficulties. Learning both to switch one's epistemological lens on a problem and to integrate different kinds of knowledge is a critical component of learning to solve problems in physics effectively. In this paper, we present two case studies in which journeyman students (upper-division physics majors) demonstrate switching between epistemological resources in approaching a complex problem. We conjecture that mas-tering these epistemological skills is an essential component of learning complex problem solving in physics.

## I. INTRODUCTION

Becoming an expert in a complex scientific field like physics means more than mastering a vocabulary, learning facts, developing skill with mathematical tools, or even developing the ability to carry out scien-tific discourse. Although these are necessary compo-nents of learning to be a scientist, the critical element of any science is creating new knowledge – learning what types of knowledge count as proof (i.e. the ways we decide we know something in a particular disci-pline). In physics, or in any scientific field that uses mathematics heavily, a variety of epistemologically distinct reasonings (that is, proof-making strategies) may be brought to bear including: mathematical calcu-lation, mapping physical meaning into mathematics, citing authoritative sources, and seeking mathematical consistency, among others.[1] As a result, developing epistemological sophistication is a critical element of becoming a professional scientist. This paper's goal is to detail what we mean by "epistemological sophisti-cation".

In previous papers[1,2,3] we have demonstrated that in all classes at the university level, students can "get stuck" in particular modes of constructing knowledge, using a limited set of the knowledge they possess while ignoring other things they know well that would have helped them significantly in making progress to a solution. We argue here that although this can occur at all levels (even with professionals), a critical part of the novice to expert transition in physics is learning to integrate different kinds of knowledge in the solution of a problem. This is particularly important at what we refer to as *the journeyman level* – that level where students have developed sufficient skills that they can no longer be considered novices but where they have not yet had sufficient experience with sophisticated problem solving (and research) to be considered ex-perts. For most students, we intend this stage to corre-spond to the upper division and early graduate level.

Typically, we expect the kind of epistemological development described here occurs as a tacit part of our upper division and graduate instruction in physics – an element of the *hidden curriculum* of upper divi-sion physics. Experts may automatically reach for a variety of epistemic tools and take their integration for granted, and upper division instructors may not realize the nature of the difficulty their students are having. We have often heard colleagues complain that their students "don't know enough math", despite having successfully passed (and even done well in) the rele-vant formal math classes. In our experience, many of these apparent "difficulties with math" are difficulties with juggling and integrating the various mathematical lenses available– epistemological difficulties of not knowing which elements of their mathematical knowl-edge to call on in particular situations.

We suggest that for many students and instructors, making these issues of epistemological sophistication explicit may be productive in creating upper division instruction that is effective for a larger fraction our physics majors.

In this paper we draw on our observational database of students working on problems in upper division physics. In an attempt to explicate this epistemological sophistication we claim is vital to expertise, we present two case studies that show journeyman students displaying it. We discuss these case studies within the theoretical framework of resources[4,5,6] and use the explicit terminology of epistemological resources, epistemological framing, and warrants elaborated in our previous paper.[1, 7]

Since this paper aims to detail this epistemological sophistication component of expertise in physics, we should briefly situate it within the (very large) body of literature about expertise in physics and problem solving. The reader looking for more detail is urged to consult section 2.3 of Reference 7 (and the references therein).

There are two broad perspectives on expertise. One is that experts behave like experts because they have larger stores of knowledge that are much more efficiently indexed than novices do. See, for example, the classic study by Larkin et al.[8] The other perspective holds that experts are better problem solvers because they are better in-the-moment navigators. That is, they are better at evaluating their current progress in real time and appropriately adjusting their thought trajectories. See, for example, Schoenfeld's account of his problem-solving class.[9] Our aim in this paper is to explicate what we see as a vital (and often underrepresented) part of expert physicists' in-the-moment navigation during problem solving: their epistemological sophistication.

In section II, we briefly recap the theoretical framework and the language we have previously developed to describe the relevant epistemological issues in students' thinking. In section III, we describe the study from which our data is drawn, discuss our methodology, and briefly review the results in our previous papers.

In section IV we present the first of our case studies, in which two students solving a problem in quantum physics admirably confront the epistemological implications of a mathematical exercise. They nicely illustrate the fluid epistemological frame-switching we conjecture is an important component of expertise in physics. In section V, we present our second case study, in which an upper-level undergraduate attempts unsuccessfully to solve a problem in vector calculus, but again demonstrates a high level of epistemological sophistication. He not only nimbly switches framings but also searches for coherence between them. In section VI, we state our conclusions and consider the implications for instruction.

## II. THEORETICAL FRAMEWORK

### A. The Resource Framework

In this work, as in our previous work, we are developing an ontology of student cognition in physics – a language to describe the elements of student thinking and reasoning in order to better understand the components of the transition from novice to journeyman to expert. To do this, we work within the *Resource Framework* (RF). This is a theoretical structure for creating phenomenological models of high-level thinking based on a combination of core results selected from educational research phenomenology, cognitive/neuroscience, and behavioral science. It is a framework rather than a theory in that it provides ontologies (classes of structural elements and the way they behave) and it permits a range of possible structures and interactions built from these elements.

The key element of the framework is that it focuses on small bits of knowledge, resources, that can be activated and associated in a variety of context dependent ways. For this paper, the critical elements of the RF are that the activation of mental networks of associated elements is *dynamic* and that there are *control structures*. The first means that the particular elements or networks of elements that are activated can change in response to new (external or internal) input. The second means that there are elements in the structure that use partial information to control which elements and are activated and which are ignored.

An extensive bibliography of papers developing and working within the RF is given on the website of the University of Maryland Physics Education Research Group.[10] (For more detail particularly relevant to this paper, see references 1, 5, and 6.)

### B. Epistemological Elements of the Framework

One important component of the RF for this work is the identification of epistemological elements of a student's knowledge structure. By *epistemology* we mean the students' knowledge about knowledge. How do they decide what knowledge is relevant to bring to bear in solving a particular problem or component of a problem? What counts as valid proof in this case? How do they decide something is known?

Note that within the RF, when we talk about student epistemologies we are not talking about their broad general beliefs about the nature of science. Rather, we are referring to their *functional epistemologies* (or "personal epistemologies")[11, 12] – how they decide they know in a particular context in a particular moment. These functional epistemologies are often highly dynamic and labile. Thus, we might ask a student "whether well-established authoritative results are important in science", and they might respond "yes" or "no" and they may even do so in a reliable and repeatable manner. But in solving a problem, a student may "choose" (consciously or unconsciously) either to simply quote a theorem or to rebuild that theorem, essentially recreating the proof, in order to see how the conditions of the theorem play out in a particular context.

These and other epistemological elements of knowledge structures in the RF have been developed in a series of papers by Hammer, Elby, Lising, Bing, and Redish.[11][12][13] The three critical concepts for us here are *warrants*, *epistemological resources*, and *epistemological framing*. The reader is encouraged to refer back to our previous paper[1] for a more detailed development of these three critical concepts.

We use the term *warrant*, borrowed from the rhetoric and argumentation theory literature,[14] to focus on the epistemic content of a statement. In rhetoric a warrant is a reason to believe, a connection between a claim and data given in support of that claim. For example, we might state that Johannes Gutenberg was the greatest inventor of the second millennium (claim) because he invented the movable-type method of printing books (data). The relevant warrant that would link this data to that claim would be that movable type allowed for the cheap mass distribution of printed material, astronomically boosting the common citizen's access to information and learning. Warrants are often tacit, a part of the communal knowledge assumed in any conversation.[*] In our study of student reasoning in upper division physics, we found that students often explicated their warrants when in a situation where there was a disagreement or a challenge. Our focus on these articulated warrants allowed us to begin to develop a classification scheme.

In our observation of many warrants articulated in our study, we found that they could be classified under a variety of general statements that we refer to as *epistemological resources*.[11,12,13] In our previous study[1] we identified four epistemological resources:

*Calculation* – algorithmically following a set of established computational steps should lead to a trustable result.

*Physical Mapping* – a mathematical symbolic representation faithfully characterizes some feature of the physical or geometric system it is intended to represent.

*Invoking Authority* – information that comes from an authoritative source can be trusted.

*Mathematical Consistency* – mathematics and mathematical manipulations have a regularity and reliability and are consistent across different situations.

We found most of the explicit warrants used by students in conflict or challenge situations fit one of these categories. This is not meant to be an exclusive list.

The third term we use in describing students' functional epistemological stances is *epistemological framing*. This refers to the assumption in the RF that there is a control process asking the question, "What is it that's going on here?"[15][16] This process certainly occurs at the beginning of any activity and, although it may result in the individual activating resources that govern an individual's behavior for an extended time period, there is a continual dynamic checking process that can lead to a reframing of the activity. For example, a medical doctor on call attending a concert may be attentive to the music for many minutes until his pager (set on vibrate) redirects his attention. Framing an activity includes many components including: social (Who will I interact with – and how?), artifacts (What materials will I use?), skills (What will I be doing here?), affect (How will I feel about what I'm doing?), and epistemology (How will I build new knowledge?). For this work, we focus on epistemological framing – the judgments that are made (often implicitly) as to what knowledge is relevant to bring to bear in a particular situation.

Note that the three terms "warrants", "epistemological resources", and "framing" are not intended to represent a strict, rigid decomposition of student activity. Rather, they represent emphases of different facets of what may appear a single unified cognitive activity, both to the individual making the statements and to the researcher.[17]

## III. THE STUDY

### A. Methodology

The study from which the data for this paper was drawn was a part of a project to observe and analyze student thinking, in particular their use of mathematics, in upper division physics.[18] We collected approximately 150 hours of video data of upper-level undergraduate physics students solving problems for such classes as Quantum Mechanics I and II, Intermediate Mechanics, Intermediate Electricity and Magnetism, and Intermediate Mathematical Methods. Data involved both students working in groups on authentic

---

[*] A warrant is indeed a function of assumed community knowledge. If, in our previous example we could assume that our listeners all accepted the idea that "boosting the common citizen's access to information and learning" was of primary importance, we are done. If that community does not all accept that idea, it becomes a claim and we have to come up with new data and a new warrant. It is the level of common communal agreement assumed that prevents the claim/data/warrant chain from iterating ad infinitum. See L. Carroll, "What the tortoise said to Achilles," *Mind* **4**(14) 278-280 (1895).

class problems and individual problem-solving interviews with problems set by the researchers.

Our identification of the four common epistemological framings in our upper level physics students' use of mathematics (i.e. Calculation, Physical Mapping, Invoking Authority, and Mathematical Consistency) originally came from this data set. We used an iterative methodology to identify these common framings, first describing the thinking of the students in a small subset of episodes and forming what generalizations we could. We then tested these generalizations against a new subset of episodes, refining where necessary. Our four common framings and their descriptions emerged after many such iterations. The reader is encouraged to consult Reference 7 for a more detailed methodological description (including details of an inter-rater reliability test for identifying these four framings).

The two case studies chosen for this present paper were chosen because they stood out from the general data set in an interesting way. In contrast to cases we have previously reported in which students appeared to "get stuck" in inappropriate epistemological frames, the students in these examples were seen to shift their epistemological framings very quickly, yet in a controlled manner (that is, their fellow students and the interviewer had little trouble following their line of thinking). These students appeared (at least in these moments) to be notably expert-like in their in-the-moment navigation of their problem solving, and they are thus held up as prototypical examples of the "epistemological sophistication" we seek to develop in our students. We chose the cases in this paper in part because the students involved did not "get the problem right". The dissociation between mathematical correctness and epistemology illustrates our point that dynamic handling of epistemological framing is an independent strength of advanced journeymen.

Intuition has many components: the identification of identity – to determine when things are supposed to represent the same thing (as in being able to follow formal proof); and the making of meaning – placing a problem in a broader context by linking to the many things we know about our subject and about the world.

## B. Recap of Previous Results: Getting stuck in an epistemological frame

In our previous work[2,3,1] we have given examples of students "getting stuck" in an epistemological framing that is inappropriate for solving the problem they are working on. The result is that for many minutes they fail to bring to bear knowledge that they both know very well and that would have shown them their current approach was in error. We briefly recap three examples here at three levels, referring the reader to the original publications for more detail.

In their observations of novice physics students working homework problems in introductory college physics, Tuminaro and Redish[2] observed a group of students solving an estimation problem near the end of the first semester. The problem asked the students to "estimate the difference in atmospheric pressure between the floor and ceiling of your dorm room." Estimation problems had been a recurrent feature in both homework and exams through the term. Yet despite the explicit terminology intended to cue an estimation framing, one student was observed leading her group through an activity that Tuminaro and Redish referred to as the "recursive plug-and-chug" epistemic game.[19] This activity is part of a framing that activates the epistemological resources of calculation and invoking authority. The student identifies a target variable (pressure) and selects an equation from the book ($PV = nRT$). The student makes explicit statements that clearly indicate that she believes that any numbers that she uses must come from authority and that, by implication, she may not construct them out of her own experience. She also makes explicit statements that clearly indicate that she does not consider the physical meaning of either the equation chosen (she fails to distinguish pressure from change in pressure) or of the symbols in the equation (she first interprets "$R$" as "the radius" and is pleased when someone says, "no, it's just a known constant"). This inappropriately restricted framing persists for many minutes and she is explicitly resistive, ignoring repeated and increasingly explicit hints from the TA that she needs to reframe the activity as an estimation problem.[20]

In a second example using data from the current study, Bing and Redish[3] observed a group of students in an undergraduate quantum mechanics class trying to evaluate an integral arising from problem 5.6 in Griffiths' text, *Introduction to Quantum Mechanics*.[21] Although the problem is about the difference between fermions and bosons, an intermediate stage requires the students to evaluate the expectation value of $x^2$ in a stationary state of the one-dimensional infinite square well. The students write down the expectation value $\int x^2 |\psi_n(x)|^2 dx$. They quickly convert this largely abstract representation into a more concrete one, $\frac{2}{L}\int_{-\infty}^{\infty} x^2 \sin^2\left(\frac{n\pi x}{L}\right) dx$, not noticing that they have incorrectly identified the limits on the integral. They then proceed to frame the task as one of Calculation. The student leading the work has many powerful tools available including her own skill with formal integration and her ability to use external calculation tools including Mathematica™ on a laptop and a symbolic integrator on a TI calculator. These tools inform her that the simplified integral she has chosen to work with, $\int_{-\infty}^{\infty} x^2 \sin^2 x dx$, is undefined. Because the set of cognitive resources she has activated do not include either her physical knowledge about the problem (or, indeed, a graphical representation of the mathematical integral), she focuses on the "algorithmically following a set of established computational steps". She becomes increasingly convinced that the result is undefined and that the problem is unanswerable. It's only when one of the students reframes the task more physically ("Hey, it's not infinity to infinity…We only have to integrate over the square well!") do they all realize (with some embarrassment) that they had just spent fifteen minutes trying to calculate the wrong integral.

Our third example of "getting stuck" is taken from the antecedent paper to this work.[1] In that paper, we discuss an example in which two students are arguing over the solution to a problem in a class in Intermediate Mathematical Methods. The task was one from vector calculus: to evaluate a change in gravitational potential energy by doing the integral of the work, $\int_{A}^{B} \vec{F} \cdot d\vec{r}$, over two distinct paths. One student recalls the theorem that for conservative forces the work is path independent and frames the task as Invoking Authority, citing the theorem without producing any further evidence despite increasing requests from the other discussant. The second discussant has framed the task as Physical Mapping, insisting that the longer path must correspond to the larger integral. The interaction is made more complicated by the fact that the first student has incorrectly written down the explication of the abstract integral in a way such that explicit evaluation supports the view of the second student. This discussion lasts for many minutes with neither student being willing either to accept the other student's warrants or to accept the other's epistemological framing. Only when they succeed in negotiating a common epistemological frame does the discussion lead to an effective solution.

These three examples all show novice and journeyman students getting stuck in a particular epistemological frame. As the main point of this paper, we want to now present two case studies that demonstrate how journeyman students are able to dynamically shift their epistemological frames. We conjecture that this fluid epistemological frame shifting is an important hallmark of expertise in physics. These examples are presented as "dissociations," as examples in which the students show skill and flexibility in epistemological frame switching but in which this does not suffice to lead them to the correct answer. These students all struggle, but they struggle in a sophisticated way. This illustrates our point that the epistemological component of expertise is independent of the correctness of a student's answer.

## IV. CASE STUDY 1: FLEXIBLE FRAMINGS

Our first case study has two physics students trying to agree on the best way to frame the math use at hand. S1 will make several framing bids. S2 responds to these bids, illustrating how epistemological framing can be a relatively labile process as well a "sticky" one. We will use our previously developed warrant analysis scheme[1] to identify the epistemological framings (Calculation, Physical Mapping, Invoking Authority, and Mathematical Consistency).

### A. The question

The two students in this episode are enrolled in a second semester undergraduate quantum mechanics class. They are meeting outside of class to work on that week's homework assignment. The case study begins with the students part way through problem 6.32, part b, in Griffiths's undergraduate quantum text.[21] That problem deals with the Feynman-Hellmann theorem, $\frac{\partial E_n}{\partial \lambda} = \langle \psi_n | \frac{\partial H}{\partial \lambda} | \psi_n \rangle$, which relates the partial derivative of an energy eigenvalue with respect to any parameter $\lambda$ to the expectation value of the same partial derivative of the Hamiltonian.

The problem tells them to consider the one-dimensional harmonic oscillator, for which the Hamiltonian is $H = \frac{-\hbar^2}{2m} \frac{\partial^2}{\partial x^2} + \frac{1}{2} m \omega^2 x^2$ and the $n^{th}$ eigenvalue is $E_n = \hbar \omega \left( n + \frac{1}{2} \right)$. They are asked to set $\lambda$ equal to $\omega$, $\hbar$, and $m$ (the angular frequency of the oscillator, Planck's constant, and the mass of the oscillator, respectively) in turn and to use the Feynman-Hellmann theorem to get expressions for the oscillator's kinetic and potential energy expectation values.

We begin with S1 noticing an oddity. When she sets $\lambda = \hbar$, the Feynman-Hellmann theorem requires her to consider $\frac{\partial}{\partial \hbar}$. How does one deal with a partial derivative with respect to a universal constant?

### B. A framing clash and a quick shift

The two students are seated at a table throughout this discussion. They do not gesture towards any diagrams or equations in a shared space.

1. **S1**: If we figure this out, hopefully it'll make the other ones easier. When you say something's a function of a certain parameter, doesn't that mean that as you change that parameter, the function changes?

2. **S2**: mmm-hmm
3. **S1**: OK, so I can change omega, but I can't change h-bar.
4. **S2**: Sure you can.
5. **S1**: I can?
6. **S2**: You can make it whatever you want it to be.
7. **S1**: But
8. **S2**: It's a constant in real life, but it's a funct-, it's, it appears in the function and you're welcome to change its value.
9. **S1**: But then it doesn't mean anything.
10. **S2**: Sure it does. Apparently it means the expectation value of [kinetic energy].
11. **S1**: You don't really know what you're talking about.
12. **S2**: Look, all it is, is you're gonna take the derivative with respect to
13. **S1**: Yeah, I understand what they want me to do here.
14. **S2**: They're just applying the theorem.

S1 begins this passage with a concise check on what a derivative entails. "When you say something's a function of a certain parameter, doesn't that mean that as you change that parameter, the function changes?" (line 1). Upon S2's affirmation, S1 points out a mismatch of this mathematical point with a physical reality. The parameter $\hbar$ is a universal physical constant. Taking a partial derivative with respect to $\hbar$ would imply that Planck's constant can vary. S1 is framing her use of mathematics as Physical Mapping. Her warrant for not accepting the $\frac{\partial}{\partial \hbar}$ operation focuses on how valid uses of math in physics class tend to align with physical reality.

S2 initially responds to S1's concern by asserting a rule. The warrant for his counterargument concerns the practical, common use of statements and previous results without explicit justification. "Sure you can [change $\hbar$]" he says. "You can make it whatever you want it to be" (lines 4 and 6). In so responding, S2 is lobbying for an Invoking Authority framing. He is suggesting S1 set aside her physically motivated objections and instead judge the validity of $\frac{\partial}{\partial \hbar}$ according to his confidence in his assertions.

S1 and S2 are arguing over something much deeper than whether or not one is allowed to take a partial derivative with respect to $\hbar$. They are disagreeing over what would be appropriate grounds for accepting or rejecting such an operation.

S1 does not accept S2's bid for Invoking Authority. Upon her first protest in line 7, S2 quickly admits "it's a constant in real life" (line 8) but sticks to his Invok-ing Authority framing. "It appears in the function and you're welcome to change its value" (line 8).

S1 protests again; "But then it doesn't mean anything" (line 9). Such a statement's full interpretation relies on acknowledging S1's Physical Mapping framing. In some framings, S1's statement is patently false. The operation $\frac{\partial H}{\partial \hbar}$ can "mean" plenty. For example, in a Calculation framing carrying out the operation on the Hamiltonian operator as a formal mathematical calculation without any consideration of physical meaning would produce the operator $\frac{-\hbar}{m}\frac{\partial^2}{\partial x^2}$. The mathematical operations involved in such a computation (and their formal interpretations and theoretical underpinnings) were among the crowning achievements of calculus's discovery. S2 retains his Invoking Authority framing and quickly responds with another "meaning" of $\frac{\partial H}{\partial \hbar}$. Quoting from the textbook's statement of the homework problem, he says "Sure it [means something]. Apparently it means the expectation value of [kinetic energy]" (line 10). The question had told them to set $\lambda = \hbar$ in the Feynman-Hellmann theorem, $\frac{\partial E_n}{\partial \lambda} = \langle \psi_n | \frac{\partial H}{\partial \lambda} | \psi_n \rangle$, and hence to obtain an expression for the expectation value of kinetic energy. S2 is thus relying on the authority of the text's question for his interpretation of $\frac{\partial H}{\partial \hbar}$. Only by acknowledging S1's current Physical Mapping framing can we place her claim in the proper context. If one's warrant for judging an operation like $\frac{\partial H}{\partial \hbar}$ concerns the alignment of the mathematics with a physical reality, then yes, that operation can be said not to "mean" much of anything. In the real physical world Planck's constant has a particular value and does not vary.

S1 objects to S2's arguments again in line 11. "You don't really know what you're talking about." This perturbation was sufficiently strong to cause S2 to reframe his attempt to justify $\frac{\partial}{\partial \hbar}$. He says "look, all it is, is you're gonna take the derivative with respect to" (line 12) before getting cut off by S1. Coupled with his next statement in line 14, "they're just applying the theorem," these statements can be seen as an attempt to reframe his thinking as Calculation. S2 is suggesting they go ahead and use their calculus machinery to take the partial derivative. As long as they stay true to the rules of calculus, they should be able to trust whatever result appears.

S1 acknowledges this attempt to reframe their work as Calculation. "Yeah, I understand what they want me to do here" (line 13). Lines 12 to 14 nicely illus-trate how efficient this implicit epistemic frame negotiation can be. These lines didn't even take five seconds to speak. In those five seconds, S2 made a call for using a different set of warrants. S1 heard that call and her brain quickly activated some of the procedures and techniques that would be associated with such a framing, as evidenced by "yeah, I understand what they want me to do here" (line 13). S2, just as quickly, acknowledges S1's acknowledgment of his reframing suggestion with his "they're just applying the theorem" (line 14).

### C. Another quick shift, this time to a shared physical mapping framing

S1 still insists on a justification more in line with her Physical Mapping framing. She begins the next chunk of transcript with another reframing objection. S2 responds by nimbly dropping his Calculation framing and adopting Physical Mapping himself.

15. **S1:** But I don't understand how you can take the derivative with respect to a constant.
16. **S2:** Because if you change the constant then the function will change.
17. **S1:** But then it's not, it's not physics.
18. **S2:** So? Actually it is, 'cause, you know, a lot of constants aren't completely determined.
19. **S1:** There's still only one value for it, that's what a constant is.
20. **S2:** The Hubble constant changes. The Hubble constant changes as we improve our understanding of the rate of expansion of the universe, and we use the Hubble constant in equations.
21. **S1:** But there's only one, right, there's only one constant. It does not vary.
22. **S2:** Yeah, but the value's changing as we approach the correct answer.
23. **S1:** It's just gonna get fixed. That's not, that's not helping us with the derivative.
24. **S2:** You can always take a derivative with respect to anything.
25. **S1:** But if you take it with respect to a constant, you'll get zero.
26. **S2:** Not if the constant itself appears in it. The derivative tells you if you change whatever you're taking the derivative with respect to how will the function change?

S1 begins this block of transcript by repeating her discomfort with $\frac{\partial}{\partial \hbar}$ (line 15). S2 responds with "because if you change the constant then the function will change" (line 16). This statement does not clearly align with only one of this paper's common framings.

Its ambiguity comes in large part from its isolation. Perhaps it was a prelude to a calculation explanation, or perhaps S2 was preparing to use some sort of Math Consistency warrant as he related this $\frac{\partial}{\partial \hbar}$ issue to a more familiar Calculus 101 example. S2's thought could have evolved this way or that, but one cannot assume line 16, by itself, was necessarily the tip of an implicit iceberg of coherence.

S1's next objection, "but it's not, it's not physics," (line 17) leads S2 to start explicitly searching for an example of a physical constant that varies. In undertaking such a search, S2 has adopted the framing S1 has been pushing. Valid use of math in physics class should align with physical reality. S2 hopes that by finding an example of a varying physical constant he can convince S1 that it is permissible to take a derivative with respect to Planck's constant. S2 frames his activity as physical mapping starting in line 18.

S2 invokes the analogy of the Hubble constant in lines 18 to 22. The Hubble constant is connected to the rate of expansion of the universe. S2 points out that the value of the Hubble constant quoted by scientists has changed over the past half a century as our measurement techniques have improved. He argues that the Hubble constant, variable as it seems, is an important part of many physics equations. By extension, it should be permissible to consider a varying Planck's constant.

S1 offers a much richer response to S2's Hubble constant argument than she has to any of his other attempts in this episode. Up to this point, she had been simply shooting down S2 with comments like "but then it doesn't mean anything" (line 9), "you don't really know what you're talking about" (line 11), and "but then it's not, it's not physics" (line 17). S2's Hubble constant argument marked the first time he adopted S1's warrant concerning the alignment of math and physics, i.e. the first time he and S1 shared a common epistemological framing.

This shared epistemological framing helps S1 engage with S2's chosen example in lines 21 to 23, and she points out that he's confusing a measurement variance with an actual physical variance. Sure, she says, our quoted value for the Hubble constant has shifted as our measurements improve, but presumably our measurements are tending towards a fixed value. The Hubble constant itself, she says, isn't changing. "That's not helping us with the derivative" (line 23).

This counterargument causes S2 to reframe the situation once again as he turns to a different type of justification. He quotes a rule again in line 24. "You can always take a derivative with respect to anything." S1 misspeaks when she replies. "But if you take it with respect to a constant, you'll get zero" (line 25). This statement seems to confuse her earlier correct interpretation of $\frac{\partial}{\partial \hbar}$ (as in line 1) with the Calculus 101 mantra "the derivative of a constant is zero", i.e. $\frac{\partial \hbar}{\partial x} = 0$. S1 responds to this misstatement in line 26.

### D. A Final Frame Shift

The final block of transcript from this episode follows S2's quick correction. It begins with S1 objecting yet again and S2 trying out yet another framing.

27. **S1**: So I don't understand how you can change a constant.
28. **S2**: You pretend like it's not a constant. It's just like when you take partial derivatives with respect to, like variables in a function of multivariables. You pretend that the variables are constant.
29. **S1**: Yeah, I don't have a problem with that.
30. **S2**: You're going the other way now. You're pretending a constant is a variable. Who cares?
31. **S1**: It doesn't make sense to me.
32. **S2**: You can easily change a variable—it's not supposed to, I don't think.
33. **S1**: OK, then I believe-
34. **S2**: I don't think, I don't think there's supposed to be any great meaning behind why we get the change h-bar. I think it just-they're like oh look, if you do it and you take its derivative and you use this equation, then all of a sudden you get some expectation of [kinetic energy], and you say whooptie-freekin-do.

S2 responds to S1's latest objection in line 27 via a Math Consistency framing. His newest argument relies on a warrant he hasn't yet tried: mathematics is a self-consistent field of knowledge, so a valid mathematical argument is one that fits in logically with other mathematical ideas.

S2 makes a common move for a Math Consistency framing. He draws an analogy in lines 28 to 30. In order to take a derivative with respect to $\hbar$, one has to "pretend" that the constant is a variable. S2 points out that taking a standard partial derivative with respect to one of the variables of a multivariable function involves "pretending" the other variables are constants. Their $\frac{\partial}{\partial \hbar}$ case, he argues, is "just like" that analogous example, except "you're going the other way now. You're pretending a constant is a variable."

In contrast to her more extended counterargument in the Hubble constant case, S1 rejects this present argument much more coarsely. "It doesn't make sense to me" (line 31). S2 has once again framed their work differently than S1's Physical Mapping. A plausible explanation is that each student's mind has activated a sufficiently different subset of their available mathematical resources, and that restricts the depth of their communication and interaction.

When S2 responds "it's not supposed to [make sense], I don't think" in line 32, he is explicitly addressing S1's Physical Mapping framing for the first time. While he had been responsive to her objections throughout this episode, he now argues with her epistemological framing directly. He states that he doesn't think an explanation of the type S1 seeks exists. S1 is possibly about to acknowledge inappropriateness of the Physical Mapping stance when she replies "OK, then I believe-" (line 33), but she gets cut off. S2 then elaborates a hybrid of Calculation and Invoking Authority that he sees as most appropriate in line 34. Mechanically take the derivative with respect to $\hbar$, following the familiar calculation algorithms, and then trust the Feynman-Hellmann theorem to relate this derivative to the oscillator's kinetic energy.

### E. Implications of the first case study

This case study illustrates how epistemological framing can be a relatively flexible process. The entire episode is essentially many iterations of S1 objecting and S2 saying, "Well, all right, how about this other type of explanation?" S1's objections serve as perturbations to S2's mental state. Many of them are of sufficient strength (or occur after he has reached a respectable closure point of his previous argument) to lead him to reframe his thinking. Each reframing results in S2 adopting a different type of warrant for judging the validity of his mathematical claim, that one should accept the operation $\frac{\partial}{\partial \hbar}$ as legitimate within physics, despite the constancy of $\hbar$.

This $\frac{\partial}{\partial \hbar}$ issue is a relatively difficult one. Ordinarily, a physical mapping frame is quite valuable in physics. Helping students understand the physical referents of their math is a common, if sometimes difficult, goal of many physics classes. But the situation here does not have such a direct physical interpretation. Instead, what the user of the theorem is being asked to do is consider an imaginary world, one where $\hbar$ can have a different value, and to see how comparing the value in our world to the one in this imaginary world can inform us. This kind of "breaking the expected frame" has led to many valuable results in physics, one of the most dramatic being the Bardeen-Cooper-Schrieffer explanation of the superconducting state. In this model, one of our strongest assumptions – the superselection rule that the number of leptons remains fixed – is freely violated and a state is created that contains different numbers of electrons with different probabilities.

That S1 and S2 were willing to engage in an exploration of the sophisticated epistemological issues that are implicitly involved with understanding how to frame the meaning of $\frac{\partial}{\partial \hbar}$ is commendable, even if the episode ends without an especially satisfying consensus. These students were indeed struggling, but they were struggling in an expert-like way when their thinking is viewed through an epistemological lens.

## V. CASE STUDY 2: FLEXIBLE FRAMINGS AND SEEKING COHERENCE

Our second example comes from a strong nontraditional student who had enrolled in our fourth-semester class in Intermediate Mathematical Methods at the beginning of the semester. This student already held an undergraduate science degree and had spent several years in the workplace before returning to the university to study for another degree. Upon attending the first several classes, he discovered that he was already familiar with most of the class's content. He decided to look for an option to place out of the class, which was technically required for his major. As part of the agreement reached, he took that semester's final exam some months after the course ended. When the student sat for the problem-solving interview from which the data below is taken, he had already taken the exam but he'd not yet seen how it was scored.

In the interview, the student was given a blank copy of one of the exam problems he had worked on a few days earlier. This problem dealt with three-dimensional vector calculus and was designed with an eye towards the analogous Continuity Equation the students would soon encounter in their Electricity and Magnetism class. It read:

> In class, we derived the integral relationship that expresses the conservation of matter of a fluid: $-\frac{d}{dt}\int_{\tau}\rho\,d\tau = \int_{\partial\tau}(\rho\vec{v})\cdot d\vec{A}$. Suppose that $\rho$ describes the concentration in a solvent of a chemical compound that can be created or destroyed by chemical reactions. Suppose also that the rate of creation (or destruction) of the compound per unit volume as a function of position at the point $\vec{r}$ at a time $t$ is given by $Q(\vec{r},t)$. The quantity $Q$ is defined to be positive when the compound is being created, negative when it is being destroyed. How would the equation above have to be modified? Explain.

One good way to begin this problem would be to do a dimensional analysis. Both the terms $-\frac{d}{dt}\int_{\tau}\rho\,d\tau$ and $\int_{\partial\tau}(\rho\vec{v})\cdot d\vec{A}$ have dimensions of amount of compound divided by time. The creation rate $Q$ is already a rate, so there shouldn't be an additional time derivative involved. Integrating $Q$ (which is a concentration as well) over the volume would give a dimensionally consistent third term for the equation: $\int_{\tau}Q\,d\tau$. What relative sign should be given to this third term? One way to find out would be to consider the case where there is a source of the chemical inside the volume (so $Q > 0$ by the problem's definition), but the amount of the chemical in the volume (i.e. $\int_{\tau}\rho\,d\tau$) is not changing in time (so $\frac{d}{dt}\int_{\tau}\rho\,d\tau = 0$). Some amount of chemical must then be flowing out of the volume, so $\int_{\partial\tau}(\rho\vec{v})\cdot d\vec{A}$ is positive. Thus, the $Q$ term can go on the left side with a positive sign:

$$\int_{\tau}Q\,d\tau - \frac{d}{dt}\int_{\tau}\rho\,d\tau = \int_{\partial\tau}(\rho\vec{v})\cdot dA \cdot$$

This problem was thus intended to require a mixture of physical and mathematical reasoning.

At the point where the discussion picks up, Student 3 (S3) has already read through the problem and copied the main equation, $-\frac{d}{dt}\int_{\tau}\rho\,d\tau = \int_{\partial\tau}(\rho\vec{v})\cdot d\vec{A}$, onto the blackboard and added a $Q$ term to the equation giving

$$Q(\vec{r},t) - \frac{d}{dt}\int_{\tau}\rho\,d\tau = \int_{\partial\tau}(\rho\vec{v})\cdot d\vec{A}$$

although he is not yet sure of that term's proper sign. He is not yet aware of the dimensional inconsistency of the way he included this $Q$ term. He has also already drawn a sketch showing an outflow of chemical from a region of space, to which he will refer in the upcoming transcript:

1. **S3**: yeah the one thing I was confused about on the exam and I continue to be confused about it now, is the sign of this here,
   *writes "+/–" in front of Q*
   like whether this is going to be a plus or a minus because, rate of creation, so if it's getting created, and then it's–Yeah, I'm not sure about this one, about this sign.
2. **Interviewer**: OK, so if, let's say you pick the positive sign
3. **S3**: Right.
4. **I**: OK?

5.  **S3**: Yeah.

6.  **I**: What does that then entail, that you could go check, try to check if it's right or wrong?

7.  **S3**: Uhhh, yeah, if it's a, if it's a positive sign then the right hand side has to increase

    *points to* $\int\limits_{\partial\tau}(\rho\vec{v})\cdot d\vec{A}$

    because something is getting sourced inside this volume. So for this to increase-

    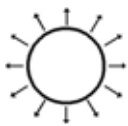

    *points to picture:*

8.  Yeah, so it cannot be a positive, it has to be a negative, because then that's going to increase- for these signs to match, for the magnitude to increase,

    *points to signs in front of* $\pm Q$ *and* $-\dfrac{d}{dt}\int\limits_{\tau}\rho\,d\tau$

    like these signs have to match,

    *Erases "$\pm$" and writes "-Q"*

    so it's probably negative.

9.  Although on the other hand, when I think of a source I think of a positive sign and sink is a negative sign. Yeah so that's where my confusion lies.

This clip begins with S3 acknowledging his confusion over the sign of the $Q$ term. Lines 1 have him putting a "$\pm$" notation next the $Q$ in his equation and noting how he wasn't particularly sure how to handle this issue several days before on the exam itself. The interviewer wanted to see how S3 would address this confusion, so he prompted S3 with a guess-and-check strategy. "Let's say you pick the positive sign…What does that entail, that you go check?" It so happens that the interviewer suggests the correct answer. $Q$ should be positive, given the side of the equation on which S3 wrote it.

The noteworthy part of this clip concerns how S3 responds to the interviewer's suggestion. S3 tries to frame the question in several different ways, trying different mathematical warrants with each framing. He doesn't disregard the previous framings' results but instead looks for consistency among the answers he gets with these different framings. His confusion on the sign of $Q$ persists, but it persists because he can't align the results his various framings. S3 (like S1 and S2) is confused, but again in an epistemologically sophisticated way.

S3 can be said to exhibit an overarching framing, one that values coherency among multiple lines of reasoning. Physical Mapping, Calculation, and Invoking Authority can be seen as subframes nested within this larger coherency-valuing framing.

S3 begins with a Physical Mapping framing in line 7. He argues that if there is a source of the chemical inside the volume (i.e. if $Q$ is positive) "then the right hand side has to increase because something is getting sources inside this volume." He had previously spent (before the quoted transcript) nearly a minute describing how $\int\limits_{\partial\tau}(\rho\vec{v})\cdot d\vec{A}$ represented a flux, an outflow of chemical from the volume. S3 is arguing that if there is a source of the chemical inside the volume, then you'd physically expect more to flow out of the volume. He juxtaposes a mathematical expression (when he points to $\int\limits_{\partial\tau}(\rho\vec{v})\cdot d\vec{A}$) with a diagram-aided physical observation of more material flowing out of the volume.

S3 makes an expert-like move when he then turns to another type of argument to hopefully support the positive-$Q$ conclusion of his Physical Mapping. His reframing is not complete. S3 is not about to simply disregard his previous reasoning in a Physical Mapping framing. He keeps his answer from the Physical Mapping framing ($Q$ should be positive) on hold to compare with what his upcoming Calculation argument will give. Line 8 has him quickly reframing the problem as Calculation. He shifts his focus to the arithmetic signs in front of the various terms in his equation: $\pm Q(\vec{r},t) - \dfrac{d}{dt}\int\limits_{\tau}\rho\,d\tau = \int\limits_{\partial\tau}(\rho\vec{v})\cdot d\vec{A}$. He notes that computationally, a positive sign in front of the $Q$ and a negative sign in front of the $\dfrac{d}{dt}\int\limits_{\tau}\rho\,d\tau$ won't have the same effect with regards to increasing the $\int\limits_{\partial\tau}(\rho\vec{v})\cdot d\vec{A}$ on the right side. "For these signs to match, for the magnitude to increase, like these signs have to match, so [$Q$] is probably negative." A negative and a negative will "match" and can work together to change the value of the right hand side.

S3's expertise does not lie in the argument he constructs in his Calculation framing. Technically, his argument is flawed. A positive $Q$ will increase the total value of the equation's left side regardless of the negative sign in front of the $\dfrac{d}{dt}\int\limits_{\tau}\rho\,d\tau$ term. S3's expertise lies in the fact that he looked to Calculation in addition to Physical Mapping. He is framing the question in different ways, nesting these framings within a larger search for coherency. Unfortunately, his two framings have produced opposite answers, so he tries a third approach.

Lines 23 to 25 have S3 reframing his work again, this time as Invoking Authority. He quickly recalls a common convention in physics (and one quoted in the problem itself). "Although on the other hand, when I think of a source I think of a positive sign and sink is a negative sign". This line of reasoning would put a positive sign in front of $Q$, contradicting the result from his Calculation framing. Still unable to find a satisfactory coherence among his arguments, S3 finishes with "yeah, so that's where my confusion lies."

This example, like the first case study, demonstrates an important component of expertise that an epistemic framing analysis can especially bring to the fore. On the one hand, S3 isn't showing much sophistication. He hasn't answered the question of the sign of $Q$, after all. On the other hand, S3 is demonstrating a very impressive component of expertise among physicists. He is approaching the problem from several different angles, trying out several different types of arguments. He is confused because he is searching for coherence among these different arguments, and he isn't finding it. Nonetheless, he is implicitly valuing this coherency. An epistemic framing analysis helps bring out this important component of his expertise.

## VI. CONCLUSIONS AND IMPLICATIONS FOR INSTRUCTION

There are two main threads of research on expertise in physics and math problem solving. First, experts have larger and better-organized banks of knowledge. Second, experts are better in-the-moment navigators during the problem solving process. Since none of the students in our case studies come to an especially satisfying final answer, it could be said that their knowledge banks are failing them.

However, the journeyman students in these case studies display a hallmark of expertise that the students stuck in particular epistemological framings in our previous papers[1,2,3] did not (at least in these given episodes). They fluidly frame their math use in different ways, either working hard to decide which is most appropriate (like S1 and S2) or looking for consistency across the different arguments their reframings produce (like S3). They display a larger, overarching framing that values this sensibility and coherence, and they can hence nest Calculation, Physical Mapping, Invoking Authority, and Math Consistency within it. The students in our previous papers display a stubborn commitment to a single frame, even in the face of reframing bids or opportunities. This epistemic framing analysis tool used here provides a lens for investigating the in-the-moment navigation component of expertise in physics problem solving. An overarching framing that values coherency among different lines of reasoning is an important component of expertise, one that can be discussed independent of the strict correctness or incorrectness of students' reasoning—an important message for both physics instructors and physics education researchers.

Our four simple epistemological resources and the associated framings that concentrate on a single one of these resources were drawn from data with novice and journeyman students. The development of expertise in physics problem solving lies well beyond any single one of these four. Expert problem solvers should have a broader, more inclusive epistemological framing in which all of these resources can be used together.

Therefore, students' reasoning should be judged by richer and more sophisticated criteria than a simple labeling of their answers as correct or incorrect or even by evaluating their mathematical skills. Framing considerations can add considerable depth to a teacher's evaluation of her student's thinking. Are the students only framing their activity in one way, or are they making an effort to approach the problem with several different framings? Are they valuing coherency among the different arguments they produce for the same problem? Even some incorrect student answers are very sophisticated from this multiple framing viewpoint, like the examples in our case studies. As teachers, we should make a special point both of explicitly modeling this search for coherency among framings for our students and seeing value when our students do it, whether they immediately get the correct answer or not.

## ACKNOWLEDGMENTS

We gratefully acknowledge the support and suggestions of many members of the University of Maryland Physics Education Research Group and of visitors to the group including David Hammer, Andrew Elby, Rachel Scherr, Rosemary Russ, Ayush Gupta, Brian Frank, Saalih Allie, and Steve Kanim. This material is based upon work supported by the US National Science Foundation under Awards No. REC 04-4 0113, DUE 05-24987, and a Graduate Research Fellowship. Any opinions, findings, and conclusions or recommendations expressed in this publication are those of the author(s) and do not necessarily reflect the views of the National Science Foundation.

## REFERENCES

[1] T. J. Bing and E. F. Redish, "Analyzing Problem Solving Using Math in Physics: Epistemological framing via warrants," *Phys. Rev. STPER*, **5**, 020108 (2009). 15 pages
[2] J. Tuminaro and E. F. Redish, "Elements of a Cognitive Model of Physics Problem Solving: Epistemic Games," *Phys. Rev. STPER*, **3**, 020101 (2007). 22 pages
[3] T.J. Bing and E. F. Redish, "Symbolic Manipulators Affect Mathematical Mindsets," *Am. J. Phys.* **76**, 418-424 (2008).
[4] D. Hammer, "Student resources for learning introductory physics," *Am. J. Phys., PER Suppl.,* **68**:7, S52-S59 (2000).

[5] E. F. Redish, "A Theoretical Framework for Physics Education Research: Modeling student thinking," in Proceedings of the International School of Physics, "Enrico Fermi" Course CLVI, E. F. Redish and M. Vicentini (eds.) (IOS Press, Amsterdam, 2004).

[6] D. Hammer, A. Elby, R. E. Scherr, & E. F. Redish, "Resources, framing, and transfer," in *Transfer of Learning: Research and Perspectives*, J. Mestre, ed. (Information Age Publishing, 2004).

[7] This paper is based on T. Bing, *An Epistemic Framing Analysis of Upper Level Physics Students' Use of Mathematics,* PhD Dissertation, University of Maryland (2008). Refer to this document for more detail. [Available at *http://www.physics.umd.edu/perg/dissertations/Bing/*]

[8] J. Larkin, J. McDermott, D. Simon, and H. Simon, "Expert and Novice Performance in Solving Physics Problems," *Science*, **208**(4450), 1335-1342 (1980).

[9] A. Schoenfeld, "Measures of problem-solving performance and problem solving instruction." In A. Schoenfeld, *Mathematical Problem Solving* (pp. 216-241). Orlando, FL: Academic Press.

[10]

*http://www.physics.umd.edu/perg/tools/ResourcesReferences.pdf*

[11] D. Hammer & A. Elby, "On the form of a personal epistemology," in B. K. Hofer, & P. R. Pintrich (Eds.), *Personal Epistemology: The Psychology of Beliefs about Knowledge and Knowing*, pp. 169-190 (Lawrence Erlbaum, 2002)

[12] D. Hammer and A. Elby, "Tapping students' epistemological resources," Journal of the Learning Sciences, 12(1), 53-91 (2003)

[13] L. Lising and A. Elby, "The impact of epistemology on learning: A case study from introductory physics," *Am. J. Phys.* **73**(4), 372-382 (2005).

[14] S. Toulmin, *The Uses of Argument* (Cambridge University Press, Cambridge, UK, 1958).

[15] I. Goffman, *Frame Analysis: An essay on the organization of experience* (Northeastern U. Press, 1997)

[16] D. Tannen, Ed., *Framing in Discourse* (Oxford University Press, 1993)

[17] It is possible that these three terms do indeed represent distinct cognitive structures, but proving this would require dissociation studies, something far beyond the scope of this work. For some methods that permit dissociations of apparently unary cognitive processes, see for example, T. Shallice, *From Neuropsychology to Mental Structure* (Cambridge U. Press, 1988).

[18] NSF grant DUE 05-24987, "Learning the language of science: Advanced math for concrete thinkers."

[19] A. Collins and W. Ferguson, "Epistemic forms and epistemic games: Structures and strategies to guide inquiry," *Educational Psychologist*, **28**(1) 25-42 (1993).

[20] This is an example of selective attention or inattentional blindness. See, for example, D. J. Simons and C. F. Chabris, "Gorillas in our midst: Sustained inattentional blindness for dynamic events," Perception, **28** (9), 1059-1074 (1999).

[21] D.J. Griffiths, *Introduction to quantum mechanics* (Pearson Prentice Hall, 2005).